\documentclass[preprint,showpacs,preprintnumbers,amsmath,amssymb,superscriptaddress]{revtex4}

\usepackage{graphicx}
\usepackage{dcolumn}
\usepackage{bm}

\begin{document}


\title{Optimization using Bose-Einstein condensation and measurement-feedback circuits}
\author{Tim Byrnes}
\affiliation{National Institute of Informatics, 2-1-2
Hitotsubashi, Chiyoda-ku, Tokyo 101-8430, Japan}
\affiliation{Institute of Industrial Science,
University of Tokyo, 4-6-1 Komaba, Meguro-ku, Tokyo 153-8505,
Japan} 

\author{Kai Yan}
\affiliation{National Institute of Informatics, 2-1-2
Hitotsubashi, Chiyoda-ku, Tokyo 101-8430, Japan}

\author{Yoshihisa Yamamoto}
\affiliation{National Institute of Informatics, 2-1-2
Hitotsubashi, Chiyoda-ku, Tokyo 101-8430, Japan}
 \affiliation{E.
L. Ginzton Laboratory, Stanford University, Stanford, CA 94305}
\date{\today}

\begin{abstract}
We investigate a computational device that harnesses the effects of Bose-Einstein condensation (BEC) to 
accelerate the speed of finding the solution of a given optimization problem. Many computationally difficult problems, including NP-complete problems, can be formulated as a ground state search problem. In a BEC, below the critical temperature, bosonic particles have a natural tendency to 
accumulate in the ground state.  Furthermore, the speed of attaining this configuration is enhanced as a 
result of final state stimulation.  We propose a physical device that incorporates these basic properties of bosons into the optimization problem,
such that an optimized solution is found by a simple cooling of the physical temperature of the device. 
We find that the speed of convergence to the ground state can be sped up by a factor of $ N$ at a given error, 
where $ N $ is the boson number per site. 
\end{abstract}

\maketitle

Quantum computation promises to offer great increases in speed over current computers due to the principle of superposition, where information can be processed in a massively parallel way \cite{nielsen00}.  The quantum indistinguishability \cite{cohentannoudji06} of particles, another fundamental principle of quantum mechanics, remains relatively unexplored in the context of information processing.  Bosonic indistinguishability is the mechanism responsible for phenomena such as Bose-Einstein condensation (BEC) \cite{pitaevskii03}. We show that by using bosonic particles it is possible to speed up the computation of a given optimization problem.
The method takes advantage of the fact that bosonic particles tend to concentrate in the minimal energy state at low temperatures. Since many difficult computational problems can be reformulated as an energy minimization problem \cite{mezard87}, this is attractive for such computational purposes that a large number of bosons lie in the ground 
state configuration. The origin of the speedup is due to bosonic final state stimulation, an effect that is familiar from stimulated emission of photons in lasers \cite{silfvast04}. This allows the system to move towards the ground state at an 
accelerated rate. 

We formulate the computational problem to be solved as an energy minimization problem of an Ising Hamiltonian \cite{mezard87}. For example, the NP-complete MAX-CUT problem \cite{ausiello99}, where the task is to group $M$  vertices into two groups A and B such as to maximize the number of connections between the groups, is known to be equivalent to the Hamiltonian $ H_P = \sum_{ij} J_{ij} \sigma_i \sigma_j $, where $ J_{ij} $ is a real symmetric matrix that specifies the connections between the sites $i,j$, and $\sigma_i = \pm 1 $ is a spin variable. The task is then to find the minimal energy spin configuration $\{ \sigma_i \} $.  In simulated annealing \cite{vanlaarhoven87}, very long annealing times are necessary to ensure that the system does not get caught in local minima. Quantum annealing \cite{das08} overcomes such problems due to local minima by introducing a quantum tunneling term but requires a slow adiabatic evolution to prevent leaks into excited states.

\begin{figure}[t]
\scalebox{0.5}{\includegraphics[bb=0 0 681 394]{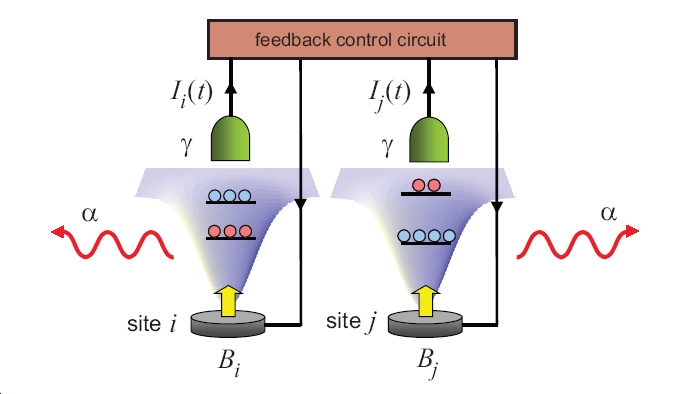}}
\caption{\label{fig1}
Each site of the Ising Hamiltonian is encoded as a trapping site, containing $N$  bosons. The bosons can occupy one of two states $\sigma=\pm1$, depicted as either red or blue. The interaction between the sites may be externally induced by measuring the average spin on each site $i$ via the detectors, which produce a detector current $I_i (t)$.  A local field on each site equal to $B_i = \Gamma \sum_j J_{ij} I_j (t) / \sqrt{\eta} $ is applied via the feedback circuit.  The system dissipates energy according to the coupling $\alpha $ to the environment.}
\end{figure}

The computational device we have in mind is shown in Figure \ref{fig1}. Each spin $\sigma_i $ in $ H_P $ is associated with a trapping site containing $N$  bosonic particles. The bosons can occupy one of two spin states, which we label by $\sigma=\pm1$. Any particle that displays bosonic statistics with an internal spin state may be used, such as exciton-polaritons in semiconductor microcavities, which have recently observed to undergo BEC \cite{deng06,kasprzak06,balili07} or neutral atoms with an unpaired
electron in atom chips \cite{folman02}.  Systems that undergo BEC are natural choices for implementation of such a device, since similar principles to the formation of a BEC are required in order for the rapid cooling to the solution of the computational problem. Exciton-polaritons possess a spin of $\sigma = \pm 1$ which can be injected by optical pumping with right or left circularly polarized laser beam. The sites are externally controlled such as to follow the Hamiltonian
\begin{equation}
\label{eq2}
H = \sum_{ij} J_{ij} S_i S_j
\end{equation}
where $ S_i = \sum_{k=1}^N \sigma_i^k $ is the total spin on each site $i$, and $J_{ij} $ is the same matrix as in $ H_P $ which specifies the computational problem. The ground state spin configuration of (\ref{eq2}) is equivalent to the original Ising model Hamiltonian $ H_P $ \cite{yan10}. This can be seen by noting that the same spectrum as $ H_P $ is obtained when the site spin  is maximized $ | S_i | = N $. The energy between these levels connect linearly as the spin on a particular site is
changed from $ S_i= -N $ to $ N $ or vice versa. 

The interaction Hamiltonian (\ref{eq2}) may be produced by measuring the total spin $ S_i $ on each site, processing those measurement results and feeding an appropriate
control signal back into the system by applying a local dc field on site $i$. For example, say at a particular instant a spin measurement of all the sites are made, giving the result $ \{ S_j \} $.  Then at that moment a local field
$ B_i = \sum_j J_{ij} S_j $ is applied on site $ i $, yielding the effective Hamiltonian $ H = \sum_{i} B_i S_i $. The measurement and the feedback process are continuous.  Although $J_{ij} $ has a large connectivity and is long-ranged, by using such a feedback method to induce the interactions there is no restriction to the kind of interactions $J_{ij} $  that can be produced in principle. The above argument can be formulated in the framework of quantum feedback control. 
We start with the Wiseman-Milburn feedback master equation \cite{wiseman93a} $ d \rho_c/d t = {\cal L}_0 \rho_c + D[C] \rho_c -i \sqrt{\eta} [ F, {\cal M}\rho_c ] + D [ F ] \rho_c $, where $ {\cal L}_0 $ is a Liouville superoperator describing the internal dynamics of the system, $ D[C] \rho =
C \rho C^\dagger - \{ C^\dagger C, \rho \}/2 $ is the Lindblad superoperator, $ C $ is the measurement operator due to the meter coupling, $ \eta $ is the detector efficiency, $ {\cal M } $ is the measurement superoperator, and $ \rho_c $ is the density matrix of the system conditional on prior measurement outcomes. We consider Markovian feedback and the system is acted on by a Hamiltonian $ H_{\mbox{\tiny fb}} (t) =  I(t) F  $, where $ I(t) $ is the feedback current due to the measurement outcome \cite{wiseman02}.

We now define each of the variables in the master equation for our specific implementation. Our system consists of 
a set of cross-coupled systems such as that shown in Fig. \ref{fig1}. First consider one particular site $ i $. 
The meter measures the $z$-component of the spin, thus we have $ C = \sqrt{\gamma} S^z_i = \sqrt{\gamma} ( - 2 n_{i -} + N )$, where $ \gamma $ is the rate constant representing the measurement strength, $ n_{i -} $ is the number operator counting the number of down spins on site $ i $, and we have assumed $ N $ bosons per site. In order that the system can dissipate energy out of the system we have a dissipation term on each site $ {\cal L}_0 \rho_c = \alpha D[S^-_i] \rho_c  $, where $ \alpha $ is a rate constant determining the time scale of the dissipation (cooling), $ S^-_i = a_{i-}^\dagger a_{i+} $, and $ a_{i\sigma} $ is the annihilation operator for a boson on site $ i $ in the state $ \sigma=\pm 1 $. The first two terms of the master equation thus describe a cooling process with a dephasing term 
originating from the measurement of the $ z $-component of the spin. The back-action of the $z$-measurement
gives a measurement superoperator $ {\cal M} \rho = S^z_i \rho + \rho S^z_i $. As a result of the feedback, on each site we apply a field in the $ z $-direction such that $ F \propto S^z_i $.

Now consider the complete feedback system as a whole. Consider applying a feedback Hamiltonian of the form $ H_{\mbox{\tiny fb}} (t) = \Gamma \sum_{i,j\ne i} S^z_i J_{ij} I_j(t) / \sqrt{\eta} $, where $ I_j(t) $ is the current resulting from the measurement of site $ j $, $ J_{ij} $ is the same matrix 
specifying the problem Hamiltonian (\ref{eq2}), and $ \Gamma $ is a overall constant. Inserting these expressions into the feedback master equation gives 
$ d \rho/d t =  \sum_i \Big[ 
\alpha D[S^-_i] \rho  + \gamma D[S^z_i] \rho + \frac{\Gamma^2}{\eta \gamma} \sum_{j\ne i} J_{ij}^2 D[ S^z_i ] \rho  - i  \sum_{j\ne i}   [ \Gamma J_{ij} S^z_i,  S^z_j \rho + \rho S^z_j ]  \Big] $.
Due to the symmetric nature of the $ J_{ij} $ matrix, the last term in the above equation can be written \cite{hofmann05} as $ -i \Gamma  [H,\rho ] $, where $ H $ is given in equation (\ref{eq2}). This gives the time evolution of the density matrix $ d \rho/d t = -i \Gamma  [H,\rho ] + \alpha  \sum_i D[S^-_i] \rho 
 + \sum_i ( \frac{\Gamma^2}{\eta \gamma} \sum_{j\ne i} J_{ij}^2 + \gamma) D[ S^z_j ] \rho $. The first term is an evolution of the system according to the Hamiltonian (\ref{eq2}), which shows that the feedback Hamiltonian $ H_{\mbox{\tiny fb}} (t) $ indeed reproduces the desired Hamiltonian (\ref{eq2}). The second term
is a cooling of the system as before, and the third is a dephasing term originating from the measurement on each site, as well as a contribution from the feedback circuit 
noise.

Initially each site is prepared with equal populations of $ \sigma = \pm1 $  spins, which can be achieved by using a linearly polarized pump laser, in the case of exciton-polaritons. The system is cooled in the presence of the interactions between the sites, by immersing the system in an external heat bath.  The readout of the computation is simply performed by measuring the total spin on each site after the system cools down by dissipating heat into the environment.  The sign of the total spin gives the information of $ \sigma_i = \pm 1$  for the original spin model.  Since the ``computation'' here is the cooling process itself, no complicated gate sequence needs to be employed to obtain the ground state.

To understand the effect of using bosons, first compare the thermal equilibrium configuration of a system described above with an equivalent system that uses classical, distinguishable particles. As a simple example, consider the two site Hamiltonian $ H = -J S_1 S_2 - \lambda N (S_1 + S_2) $, where the second term is included such as there is a unique ground state in spite of the $ S_i \leftrightarrow -S_i $ symmetry of the first term in the Hamiltonian. For a single spin on each site and $J,\lambda>0 $, the ground state configuration is $\sigma_1 = 1 $, $\sigma_2 = 1$, which we regard as the "solution" of the computational problem. We neglect the presence of an on-site particle interaction $ \propto S_i^2 $ here since we assume that the strength of the interactions $J$ produced by the induced feedback method can be made much larger than such a term which may occur naturally.

\begin{figure}[t]
\scalebox{0.5}{\includegraphics[bb=0 0 691 403]{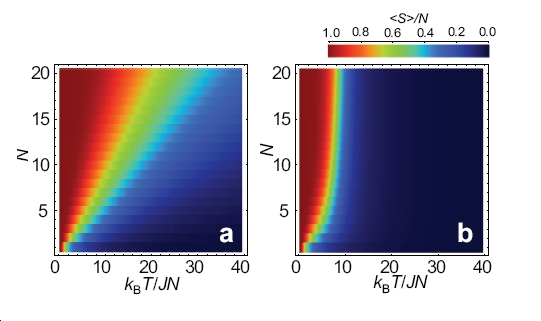}}
\caption{\label{fig2} 
The average spin of the two site Ising Hamiltonian as a function of the boson number $N$  and rescaled temperature $k_B T / JN $. (a) indistinguishable bosons and (b) classical distinguishable particles. The parameters used are $J=10$ and $ \lambda =0.5 $.}
\end{figure}

In Figure 2 we show the average spin on a single site of the two site Hamiltonian, which can be calculated from standard partition function methods accounting for bosonic counting factors. Comparing bosonic particles and classical distinguishable particles, we see that the bosonic case has a larger average spin for $N>1 $ and all temperatures, corresponding to a spin configuration closer to the ground state.  As the particle number is increased, the temperature required to reach a particular $ \langle S_i \rangle $ increases. For the bosonic case, the required temperature increases linearly with $N$, while for distinguishable particles it behaves as a constant for large $N$.  This results in an improved signal to noise ratio for the bosons in comparison to distinguishable particles. The concentration of particles in the ground state configuration for bosons is precisely the same effect that is responsible for the formation of a BEC. Since the ground state corresponds to the solution of the computational problem, this corresponds to an enhanced probability of obtaining the correct answer at thermal equilibrium.

\begin{figure}[t]
\scalebox{0.5}{\includegraphics[bb=0 0 460 419]{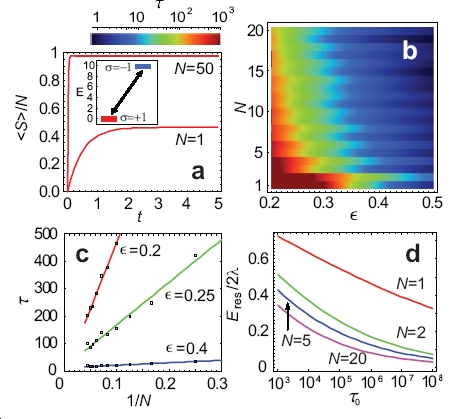}}
\caption{\label{fig3}
(a) Equilibration of a two level system with energy levels  $E_1 = 0 $ and $E_2 = 10$ at $ k_B T = 10 $
and boson numbers as shown.  (b) The equilibration time for the 4 site Ising model with $ J_{ij} = -10 $ and $ \lambda = -1 $. (c) The $1/N $ dependence of the 
4 site Ising model in (b). (d) The residual energy for the 2 site Ising model with $J=10$  and $\lambda = 0.5 $ after annealing the system with an exponential schedule with time constant $ \tau_0 $ for various boson numbers as shown. All calculations use $ \alpha = 1 $ and $ \xi = 0.001 $.}
\end{figure}

We now turn to the time taken to reach thermal equilibrium, after initially preparing the system with equal populations of $\sigma = \pm 1 $  particles on each site. We generalize the methods of Ref. \cite{glauber63} to our bosonic Ising model, also accounting for transitions beyond first order in perturbation theory. 
Given the $ M $-site Hamiltonian $ H = \sum_{ij} J_{ij} S_i S_j + \lambda N \sum_i S_i $, the states are labeled $
| \bm{k} \rangle = \prod_{i=1}^M \frac{1}{\sqrt{k_i ! (N-k_i)! }} 
(a_{i+}^\dagger)^{k_i} (a_{i-}^\dagger)^{N-k_i} | 0 \rangle  $, where the $ k_i $ range from 0 to $ N $, $ a_{i \sigma}^\dagger $ is the creation operator for a boson on site $ i $ in the state $ \sigma $, and we have defined the vector $ \bm{k} = (k_1, k_2, \dots , k_M) $. The 
probability distribution then evolves according to 
\begin{align}
\frac{d p_{\bm{k}}}{dt} & = \sum_{i=1}^M \sum_{\delta k_i = -k_i}^{N-k_i}
- w( \bm{k}, \delta  \bm{k}_i) p_{\bm{k}} 
+ w( \bm{k}+ \delta  \bm{k}_i, - \delta \bm{k}_i) p_{\bm{k}+ \delta \bm{k}_i} \nonumber
\end{align}
where $ \delta  \bm{k}_i =(0,\dots,0,\delta k_i,0,\dots,0) $ is a vector in the direction of the $i$th axis of the $ M $-dimensional hypercube.  The $ w( \bm{k}, \delta \bm{k}_i) $ is a weight factor for the process $ | \bm{k} \rangle \rightarrow | \bm{k} + \delta  \bm{k}_i \rangle $, containing
a transition rate factor from Fermi's golden rule and a
$ (1 \pm \gamma ) $ factor to ensure that the system evolves to the correct thermal equilibrium distribution, in a similar way to that discussed in Ref. \cite{glauber63}. We calculate the weight factors to have the form \cite{yan10}
\begin{align}
w(\bm{k}, \delta \bm{k}_i) = (1 + \gamma_i (\delta k) ) \frac{\alpha \xi^{\delta k -1}}{((\delta k -1)!)^2}
{\cal F} (k_i,\delta k_i) 
\end{align}
where $ \alpha $ is a rate constant determining the overall timescale,  
$ \gamma_i (\delta k) = \tanh [ 
- \delta k \beta \left( \lambda N + \sum_{j\ne i} J_{ij}(  2 k_j- N ) \right) ] $, and 
\begin{align}
{\cal F} (k, \delta k) = \left\{ 
\begin{array}{ll}
\prod_{m=1}^{\delta k} (k +m)(N-k-\delta k +m) & \delta k > 0 \\
\prod_{m=1}^{|\delta k|} (k -|\delta k| +m)(N-k +m) & \delta k < 0 
\end{array}
\right. 
\nonumber
\end{align}
The $ {\cal F} (k, \delta k)$ factors are final state stimulation factors due to bosonic statistics, from matrix elements $ |\langle k+\delta k | V^{\delta k} | k \rangle |^2 $ in Fermi's golden rule, where the perturbation causing the transition is $ V \propto a^\dagger_+ a_- + a^\dagger_- a_+ $  \cite{faisal76}. Transition beyond order one are suppressed by the coefficient $ \xi \ll 1 $.  

We use the standard numerical differential
equation solver supplied by Mathematica to evolve $ p_{\bm{k}} $ for small boson numbers. Figure \ref{fig3}a shows the cooling of the system for $N=1$ and $N=50$ particles. As the number of particles is increased, we see that the time taken to reach equilibrium is considerably reduced, as well as a high proportion of particles occupying the ground state.  For low temperatures, the time dependence of the single site case can be approximated by the rate equations $ \frac{d n_1 }{dt} =  - \frac{d n_2 }{dt}=\alpha( n_1 + 1) n_2   $, where $ n_i $ are the populations on levels $ i = 1,2$. Analytically solving this gives a equilibration time of  $ \tau \sim 1/\alpha N  $ for large $ N $, explicitly showing the final state stimulation effect. 

As is well known from past studies of simulated annealing, the presence of local minima slows down the time for equilibration dramatically. To illustrate the behavior of the system in the presence of local minima, we choose a typical four site Ising model with a local minimum state $ \downarrow \downarrow \downarrow \downarrow $ and a global minimum state $ \uparrow \uparrow \uparrow \uparrow  $. We define the error probability $\epsilon $ as the probability of failing to obtain the correct ground state configuration after a single measurement of the total spin: $ \epsilon = 1 - \sum \exp[-H/k_B T]/Z $, where $ Z $ is the partition function and the summation is over all configurations with the same sign 
of spin as the correct ground state. The effect of the local minimum can be seen from the $N = 1 $ case shown in Figure \ref{fig3}b, where the time rapidly increases as $\epsilon \rightarrow 0 $ (i.e. $ T \rightarrow 0 $) unlike the single site case. In our simulations, we assume that the Hamiltonian (\ref{eq2}) is correctly implemented by the feedback scheme, and use a kinetic Monte Carlo method \cite{voter05} to numerically calculate the cooling time starting from a $ T = \infty $ configuration. A final thermal equilibrium 
temperature is set, which determines the error probability.  Fig. \ref{fig3}b shows that as the boson number is increased, there is a significant speedup at constant error of several orders of magnitude. There is a small odd/even effect due to the definition of the error $ \epsilon $. The curves approach zero equilibriation time as 
$ \propto 1/N $ for large $ N $ (Fig. \ref{fig3}c). In all our numerical simulations we have found that bosons are able to speed up the equilibration times, with a $ \propto 1/N $ speedup for large $ N $, in a similar way to the 
single site example.

The scheme is also compatible with a thermal annealing procedure, where the temperature is gradually reduced to zero starting from a high temperature configuration. We calculate the residual energy, defined as the average energy above the ground state of the system following the annealing procedure. An exponential annealing schedule with time constant $ \tau_0 $ is used, starting from a temperature corresponding to an error of $ \epsilon = 0.7 $. Times up to $ 4 \tau_0 $ are annealed where the system no longer responds to the cooling.  Fig. \ref{fig3}d shows that the residual energy is suppressed for all $ N>1 $, thus again displaying an improvement due to bosonic final state stimulation.

We conclude that the scheme as shown in Figure \ref{fig1} has a systematic way of improving the standard Ising model, in terms of a speedup proportional to the number of bosons $ N $ per site. The origin of the speedup can be understood in the following simple way. The use of many bosons increases the energy scale of the Hamiltonian from $ \sim J_{ij} $ to $ \sim N J_{ij} $. Due to 
bosonic statistics, the coupling of the spins to the environment is increased by a factor of $ \sim N $. Thus 
by constructing a system out of bosons we have increased the energy scale of the entire problem by a factor of $ N $, which results in a speedup of $ N $. Spin flips due to random thermal fluctuations also occur on a timescale that is faster by a 
factor of $ N $, resulting in a faster escape time out of local minima.
We emphasize that although the device discussed in this letter is a computational device that uses quantum effects, it is rather different to a quantum computer, since the off-diagonal density matrix elements of the state of the device are explicitly zero at all times.   For these reasons we expect the scaling of the equilibration time with the site number $M$ is not faster than exponential, in analogy to the classical case.  The speedup then manifests itself as a suppressed prefactor of this exponential function, which can be accelerated by a factor of $ N $.  
In its present form, the device can simulate any kind of optimization problem that can be 
written as an Ising model involving two spins, such as the graph partitioning problem, 2SAT, MAX-2SAT, 
and others. Extension of the device to involve $k$-body interactions give a natural implementation 
of problems such as $k$-SAT ($k \geq 3 $).

This work is supported by the Special Coordination Funds for Promoting Science and Technology, Navy/SPAWAR Grant N66001-09-1-2024, and MEXT.  T. B. would like to thank M. Sarovar and M. Takeoka for discussions.


\end{document}